# Bending elasticity of a curved amphiphilic film decorated anchored copolymers: a small angle neutron scattering study.

**Jacqueline APPELL\*, Christian LIGOURE and Grégoire PORTE**

Groupe de Dynamique des Phases Condensées , UMR5581 CNRS-Université Montpellier II C.C.26, F-34095 Montpellier Cedex 05, France

Abstract

Microemulsion droplets (oil in water stabilized by a surfactant film) are progressively decorated with increasing amounts of poly ethylene- oxide (PEO) chains anchored in the film by the short aliphatic chain grafted at one end of the PEO chain . The evolution of the bending elasticity of the surfactant film with increasing decoration is deduced from the evolution in size and polydispersity of the droplets as reflected by small angle neutron scattering. The optimum curvature radius decreases while the bending rigidity modulus remains practically constant. The experimental results compare well with the predictions of a model developed for the bending properties of a curved film decorated by non-adsorbing polymer chains, which takes into account, the finite curvature of the film and the free diffusion of the chains on the film.

\* To whom correspondence should be adressed : appell@gdpc.univ-montp2.fr

**INTRODUCTION**

The curvature elasticity of thin liquid films has been investigated theoretically, as reviewed by Safran[1]. This curvature plays a major role in the supermolecular structures formed by self assembling molecules such as surfactants, copolymers etc... Such supermolecular assemblies are important in biological issues for i.e. the description of the flexible membranes surrounding cells. These supermolecular structures are also important in numerous applications such as cosmetics, paints, detergency etc.. The interactions between polymers and self assembled membranes where the thin liquid film is a molecular bilayer and in particular the influence of the polymers on the structure and elastic properties of the membrane are investigated by many groups see for example ref [2-5]. Our aim in this paper is to present and discuss results on the curvature elasticity of a self-assembled monolayer of surfactants and on its modification upon decoration by a polymeric chain; modification for which little information is available.

To access a monolayer surfactant film, the starting situation is a microemulsion[6]: a dispersion of oil droplets in water stabilized by a monolayer of surfactant molecules (see fig 1). The leading term of the free energy of such a microemulsion has been shown[7-11] to be the bending energy of the surfactant film. Thus the elastic properties of the surfactant film monitor the size and polydispersity of the droplets. These can be measured by small angle neutron scattering (SANS) and the elastic properties of the surfactant film can be derived. Using this approch Gradzielski et al[12-14] probed the evolution of the elastic properties of the film with the length of the aliphatic chain of the surfactant and more recently Farago and Gradzielski[15] studied the influence of the charge density on the elastic properties of a surfactant film. We use here the same approch to follow the evolution of the elastic properties of a surfactant film upon its decoration by a polymer anchored to the monolayer by its hydrophobically modified end (see fig 1).

If the microemulsion is prepared very close to the emulsification failure boundary,[9,16,17] that is the limit of oil solubilization, then the droplets of microemulsion have a radius corresponding to the optimum curvature of the surfactant film and a narrow size distribution. Qualitatively we expect this optimum curvature to increase upon decoration by polymers. Furthermore the surfactant monolayer has a constant (or almost constant) area in the sample, thus the smaller droplets form to comply with the increasing optimum curvature will be more numerous but will enclose a smaller volume of oil so that some oil must be rejected as is indeed observed. The SANS study indicates that up to a rather large amount of polymer, the droplets of microemulsion remain spherical. In fact decoration of the film by the polymer lowers the emulsification failure limit. In the present study our aim is to quantify the variations of the elastic properties of the surfactant film upon decoration by polymers.

The theoretical models developed for flat fluid membranes decorated by a small amount of anchored polymers [18-20] predict a sharper decrease of the optimum radius of curvature than that observed in our experimental results and a quite strong increase of the effective bending modulus which is not observed in our experiments Possible reasons of this discrepancies are listed below :

i/ The contribution of an ideal anchored polymer chain to both the elastic constants and the spontaneous curvature of the compound system have been calculated in the limit where the cuvature radiius of the film is much larger than the radius of gyration of the polymer chain. This approach has been successfully used to explain quantitatively the solubilization efficiency boosting by amphiphilic copolymers in *bicontinous* o/w microemulsions[21]. However this assumption breaks when applied to strongly curved interfaces, as in our experiments where the radius of the droplets is of the same order of magnitude than the polymer coil size.

ii/ In the same theoretical papers[19,20] the elastic characteristics of the surfactant film have been calculated at fixed composition of polymer chains $\sigma$. However in fluids films, anchored polymers are free to diffuse on the surface of the monolayer . Even more, in real experimental situations, they can exchange with the aqueous solvent reservoir. So, $\sigma$ is not a conserved quantity, and the mechanical properties of the mixed monolayer at fixed chemical potential $\mu$ (where $\mu$ is the potential conjugate to $\sigma$) are more relevant in this context[22,23].

Therefore we have developed a simple model ,which explicitely contain both above-mentionned ingredients to describe the evolution of the elastic properties of the surfactant film with an increasing decoration by non-adsorbing anchored polymers .

In a first section we recall how the size and polydispersity of the microemulsion droplets are related to the optimum curvature, the bending rigidity modulus of the surfactant film , the entropy of mixing of the microemulsion droplets and to their deformation. We then describe our model which gives the elastic properties of the decorated surfactant film as a function of its decoration. In the second section we describe the samples and the experimental determination of the geometrical parameters of a microemulsion through SANS measurements. In the third and last section the elastic properties of the surfactant film are derived from the experimental results and discussed, the bending rigidity modulus is not affected by the decoration of the surfactant film while the optimum curvature radius decreases regularly . These general trends are well accounted for by the proposed theoretical description.

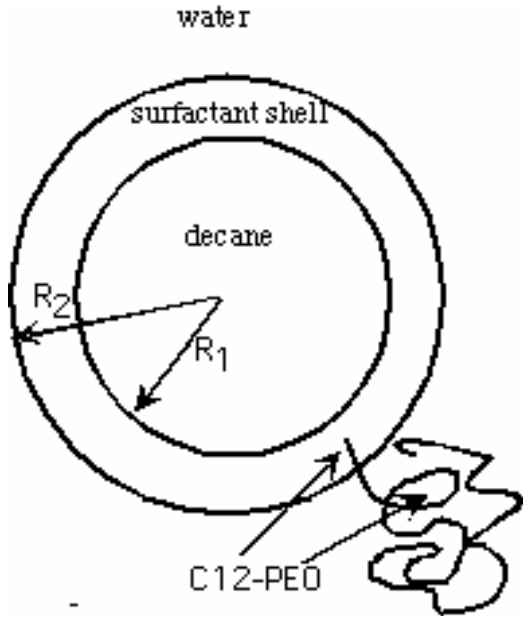

Figure 1 A microemulsion droplet with one PEO-C12 polymer anchored via its C12 aliphatic end in the surfactant shell. Water and decane are deuterated , neutron scattering is essentially due to the surfactant shell , the PEO chains are very few and do not contribute significantly to the scattering.. In the model, the shell is characterized by its mean radius $R_m = (R_1 + R_2)/2$ and its thickness $\delta = R_2 - R_1$ For the bare microémulsion $R_m \approx 61 Å$ and $\delta \approx 13 Å$ ; the PEO chain has a giration radius $R_g \approx 23.5 Å$

**Theoretical description:**

The elastic properties of the surfactant monolayer is the main ingredient in the stabilization of a microemulsion[7-10]. The free energy of the microemulsion is written in terms of the bending energy of the interface and of the entropy as[24-26]

$$F = \int dA \left[ \frac{\kappa}{2}(C_1 + C_2 - 2C_0)^2 + \overline{\kappa} C_1 C_2 \right] + NkTf(\phi)$$ with $\kappa, \overline{\kappa}$ respectively the mean

and Gaussian bending modulus , $C_0$ the spontaneous curvature and $C_1$ and $C_2$ the principal curvatures of the surfactant monolayer. These parameters depend on the composition of the surfactant layer (nature and proportion of the surfactants see e.g. ref [26] ) . For a droplet microemulsion ( $C = C_1 = C_2$ ) this can be written[17] in terms of $\tilde{\kappa} = \frac{1}{2}(2\kappa + \overline{\kappa})$ and $\tilde{C}_0 = \frac{\kappa C_0}{\tilde{\kappa}}$

$$F = \int dA \left[ \frac{\tilde{\kappa}}{2}(2C - 2\tilde{C}_0)^2 \right] + NkTf(\phi) \qquad (1)$$

the second term on the right hand side is the entropy of mixing of droplets number $N$ and volume fraction $\Phi$, we assume for $f(\Phi)$ the form proposed by Milner and Safran[27] (random mixing approximation) :

$$f(\Phi) = \frac{1}{\Phi}[\Phi \ln \Phi + (1-\Phi)\ln \Phi] \tag{2}$$

Minimizing the total free energy $F$ given by eq(1) with respect to $C$ we obtain the optimum curvature radius $R$ of the droplets as a function of $\tilde{\kappa}$, $\tilde{C}_0$ and $\Phi$

$$R\tilde{C}_0 = 1 + \frac{kT}{8\pi\tilde{\kappa}} f(\Phi) \tag{3}$$

The size and shape of the droplets are subject to thermal fluctuations[27, 28]. The form of the fluctuating droplets are described in terms of spherical harmonics and the zero order deformation is the droplet size fluctuation and can be related to the polydispersity $p$ ($p^2 = \frac{\overline{R^2} - \overline{R}^2}{\overline{R}^2}$):

$$p^2 = \frac{kT}{16\pi\tilde{\kappa} + 2kTf(\Phi)} \tag{4}$$

We neglect the higher order terms of the development. Gradzielski et al[13] have discussed their influence on the measured values of $p^2$ and justified the use of eq (4) to describe the apparent polydispersity deduced from the experimental spectra.

With eq (3) and (4) we can derive $\tilde{\kappa}$, $\tilde{C}_0$ from the measured values of $R_m$, $\delta$, $\Delta R$

$\tilde{\kappa}$, $\tilde{C}_0$ depend solely, in the bare microemulsion, on the nature and composition of the surfactant monolayer. Our objective is now to describe theoretically how they vary upon decoration of the surfactant film by the polymer in order to compare the theoretical prediction to the experimental values.

**Spontaneous curvature and bending modulus of a microemulsion of hairy droplets**

Consider a spherical droplet of radius $R = 1/C$, decorated by grafted *ideal* chains of polymerization index $N$. The number of grafted chains per unit area is denoted $\sigma$. $R_g = a\sqrt{N}/\sqrt{6}$ is the radius of giration of the ideal chain, where $a$ is the monomer length. The partition function of a grafted chain is[19]:

$$Z_{pol} = 6^N \frac{\alpha + aC}{1 + aC} \quad \text{with } \alpha = \text{erf}[a/(2R_G)] \tag{5}$$

Removing the useless constant terms, the conformational free energy of a chain is then:

$$F_{pol} = -kT \ln\left(\frac{\alpha + aC}{1 + aC}\right) \tag{6}$$

At low polymer coverage, the anchored polymers form isolated mushrooms on the droplet[22]. The mushroom regime extends up to the overlap coverage $\sigma^*$, at which the droplet becomes completely covered by anchored polymers. For flat surfaces, $\sigma_{flat}^* = (\pi R_G^2)^{-1}$. However, in the relevant experimental situation of consideration, $\sigma^* > \sigma_{flat}^*$, since $R$ and $R_g$ are of the same order of magnitude. The free energy per unit area of the hairy microemulsion droplets in the mushroom regime reads:

$$f_s = 2\tilde{\kappa}_{bare}(C - \tilde{C}_{0,bare})^2 - \sigma kT \ln\left(\frac{\alpha + aC}{1 + aC}\right) + kT\sigma \ln(\sigma a^2) \qquad (7)$$

where $\tilde{\kappa}_{bare}$ and $\tilde{C}_{0,bare}$ designe the elastic parameters of the bare droplet. The first term in equation (7) accounts for the bending energy of the bare surfactant film. The second term accounts for the addtive contibution of the conformational energy of each grafted polymer chain $F_{pol}$ ( see equ. 7). The third term is the translational entropic contribution of the dilute 2D gas of polymer anchored in the film..

Notice that we did not expand $F_{pol}$ in power of $C$, around the flat surface to obtain the elastic constants of the film, as usually done for this type of calculation [19]. Indeed since our samples belongs to the regime where $\sqrt{\pi}R_G C \approx 1$, this method of calculation would not give reliable predictions. Minimization of $f_s$ with respect to $C$ gives the spontaneous curvature of the hairy microemulsion $\tilde{C}_{O,\sigma}$ at constant polymer area density $\sigma$. It is the the single real solution of the following cubic algebric equation:

$$\frac{\partial f_s}{\partial C} = 4\tilde{\kappa}(C - \tilde{C}_{0,bare}) + \sigma kTa\left(\frac{1}{(1+aC)} - \frac{1}{(\alpha + aC)}\right) = 0 \qquad (8)$$

The analytical expression of $\tilde{C}_{O,\sigma}$ can be easily obtained, but is much too complex to be useful and is therefore not shown.

The contribution of the polymer corona to the bending modulus at constant polymer area density $\sigma$ is given by :

$$\Delta\tilde{\kappa}_\sigma = \tilde{\kappa}_\sigma - \tilde{\kappa}_{bare} = \frac{1}{4}\left.\frac{\partial^2 f_s}{\partial C^2}\right|_{C=\tilde{C}_{0,\sigma}} - \tilde{\kappa} \qquad (9)$$

with $\quad \dfrac{\partial^2 f_s}{\partial C^2} = 4\tilde{\kappa} + \sigma kTa^2\left(\dfrac{1}{(1+aC)^2} - \dfrac{1}{(\alpha + aC)^2}\right) \qquad (10)$

As shown in equation (8), the polymer composition of the membrane $\sigma$ couples to the curvature $C$, this coupling will play a significant role in the bending properties of the film, as the anchored polymers are free to diffuse in the membrane [23,29]. It will decrease the spontaneaous curvature $\tilde{C}_0$ and soften the bending modulus $\tilde{\kappa}$. We have shown in ref [29], that this properties can be interpreted in terms of a spontaneous curvature and a bending moduli calculated at fixed chemical potential $\mu$ rather than at fixed composition $\sigma$. The chemical potential $\mu$ is the potential conjugate to $\sigma$:

$$\mu = \frac{\partial f_s}{\partial \sigma} = kT \left\{ 1 - \ln(\sigma a^2) - \ln\left(\frac{\alpha + aC}{1 + aC}\right) \right\} \qquad (11)$$

We calculate now the bending modulus $\tilde{\kappa}_\mu$ and the spontaneous curvature $\tilde{C}_{0,\mu}$ at constant chemical potential $\mu$ using the general theory built up in ref [29]:

$$\tilde{\kappa}_\mu = \tilde{\kappa}_\sigma - \left(\frac{\partial \mu}{\partial C}\right)^2 \Big/ \frac{\partial \mu}{\partial \sigma}\Big|_{C=C_{0,\sigma}} \qquad (12\text{-a})$$

$$\tilde{C}_{0,\mu} = \frac{\tilde{\kappa}_\sigma}{\tilde{\kappa}_\mu} \tilde{C}_{0,\sigma} \qquad (12\text{-b})$$

with $\frac{\partial \mu}{\partial C} = \frac{kTa(\alpha - 1)}{(\alpha + aC)(1 + aC)}$, $\frac{\partial \mu}{\partial \sigma} = kT/\sigma$, and $\tilde{C}_{0,\sigma}$ (resp. $\tilde{\kappa}_\sigma$) given by equation (8) (resp. 9).

Equations 12-a and 12-b will be used to interpret our experimental results in the Results and Discussion section.

In principle the model developed above does not apply to *real* chains which are self avoiding in a good solvent. However, Auth and Gompper [30] have shown, using Monte Carlo simulations, that the universal amplitude of the bending rigidity for a self-avoiding anchored polymer does not differ too much from the result for an ideal chain. This result is consistent with the calculation by Bickel et Marques[31] of the pressure field induce by a self-avoiding chain on a flat surface. We thus expect our model to apply also for real chains, providing a suitable choice of the effective monomer length *a*.

## Experimental Section

### Materials

Cetyl pyridinium chloride [$H_3C$-$(CH_2)_{15}$] -$C_5H_5N^+$ $Cl^-$ (CPCl) from Fluka is purified by successive recristallization in water and in acetone, octanol [$H_3C$-$(CH_2)_7$]-OH from Fluka and deuterated decane [$D_3C$-$(CD_2)_8$ $CD_3$] from Sigma Aldrich are used as received. The poly (ethylene-oxide) has been hydrophobically modified and purified in the laboratory using the method described in ref [32]. After modification, the degree of substitution of the hydroxyl group was determined by NMR[32]. The degree of substitution is found to be equal or larger than 98%. The PEO-C12 polymer has a molecular weight M ~ 5200 dalton.

**Samples**

All samples are prepared by weight in 0.2M-NaCl deuterated brine. The volume fraction of the droplets in the initial microemulsion is 3.1 %. In order to adjust the optimal curvature of the surfactant layer a mixture of CPCl and octanol is used; the weight ratio of octanol to CPCl is 0.25 ; deuterated-decane is added up to the maximum solubilization so that the sample is at the emulsification failure limit, the weight ratio of decane to surfactant is 0.925 (the volume fraction of decane, in the sample, is 1.6 %). For the samples containing PEO-C12, the quantity of surfactant (CPCl and octanol with a ratio in weight of 0.25 octanol to CPCl) is adjusted upon addition of PEO-C12 to maintain constant the weight of the aliphatic part of the surfactant layer which incorporates the aliphatic chain of the copolymer. Thus the volume fraction of the shell is constant $\Phi = 0.0153$ with a variation of less than 1% between the bare microemulsion and the most decorated microemulsion. All the samples contains the same amount of deuterated decane but only part of it is solubilized as described later. Samples are characterized by the number $\sigma$ of copolymer per surfactant layer area (see below the discussion section).

The theoretical description assumes that the anchored chains are not absorbed onto the surrface ; indeed we checked earlier that PEO does not adsorb onto the surfactant film [33].

**Small Angle Neutron Scattering : SANS Measurements**

They have been performed at LLB-Saclay on the spectrometer PACE. The range of scattering vectors covered is 0.004 Å$^{-1}$ < $q$ < 0.16 Å$^{-1}$ . The temperature is T= 20°C. The scattering data are treated according to standard procedures. They are put on an absolute scale by using water as standard. And we obtain intensities in absolute units (cm$^{-1}$) with an accuracy better than 10%. To simulate correctly the experimental spectra all the model spectra are convoluted by the instrumental response function taking into account the actual distribution of the neutrons wavelength and the angular definition[34].

To gain in accuracy in the determination of the size and polydispersity of the microemulsion droplets we made experiments under shell contrast conditions: the droplets of deuterated decane are surrounded by an hydrogenated surfactant film and dispersed in deuterated water or brine. Deuterated decane and deuterated water have almost equal scattering length density (6.6 10$^{10}$ cm$^{-2}$ and 6.4 10$^{10}$ cm$^{-2}$) so that a droplet is viewed as a spherical shell formed by the surfactant layer. We neglect in our interpretation the very small perturbation brought about by the PEO chains on the scattering length of the solvent surrounding the droplets.

The small angle neutron scattering from colloidal solutions provides information on their structure. If the colloidal aggregates can be assumed to be spherical or if, at least on average, the interaction potential between them has spherical symmetry, one can write the scattered intensity $I(q)$ ($cm^{-1}$) in the form:

$$I(q) = \Phi v (\Delta \rho)^2 P(q) S(q) = A P(q) S(q)$$
$$\text{with } A = \Phi v (\Delta \rho)^2 \qquad (13)$$

where $q$ ($Å^{-1}$) is the scattering vector; $\Phi$ is the volume fraction of aggregates (here the surfactants shell); $v$ ($cm^3$) the dry volume of the aggregates and $\Delta\rho$ ($cm^{-2}$) the contrast i.e. the difference in the scattering length density of the aggregates and of the solvent. $P(q)$ is the form factor of the colloidal aggregates and $P(q \to 0) = 1$. $S(q)$ is the structure factor which reflects interactions between the aggregates, at large values of $q$, $S(q) \to 1$. We will focus on the range of $q$ where $S \approx 1$.

For a model of concentric shells with sharp boundaries (cf figure 1). We write the form factor $P(q)$ as a function of $R_m = (R_1 + R_2)/2$ and $\delta = R_2 - R_1$

$$P_{shell}(q) = \frac{16\pi^2}{v_{shell}^2 q^6} \left[ 2qR_m \sin(qR_m) \sin(q\delta/2) + 2\sin(q\delta/2) \cos(qR_m) - q\delta \cos(qR_m) \cos(q\delta/2) \right]^2$$

$$\text{and } v_{shell} = \frac{4\pi}{3} \delta (3R_m^2 + \frac{\delta^2}{4}) \qquad (14)$$

If $\delta \ll R_m$ and $q\delta < 1$ eq. 14 reduces to $P_{shell}(q) = \sin^2(qR_m)/(qR_m)^2$. We use a $q^2 I(q)$ representation which amplifies the oscillations of $P(q)$; in a first approximation the extrema are those of the function $\sin^2(qR_m)$.

The model spectra have been calculated using eq. 14 for $P(q)$. The polydispersity of the droplets size is described by a Gaussian distribution [35] of the radius of the droplets. In the shell contrast we assume a constant shell thickness $\delta$, a mean value $\overline{R_m}$ and a standard deviation $\Delta R$.

$$AP(q) = \Phi_{shell} (\Delta\rho)^2 \int \frac{v}{\sqrt{2\pi}\Delta R} P_{shell}(q, R_m) . e^{-\frac{(R_m - \overline{R_m})^2}{2(\Delta R)^2}} dR \qquad (15)$$

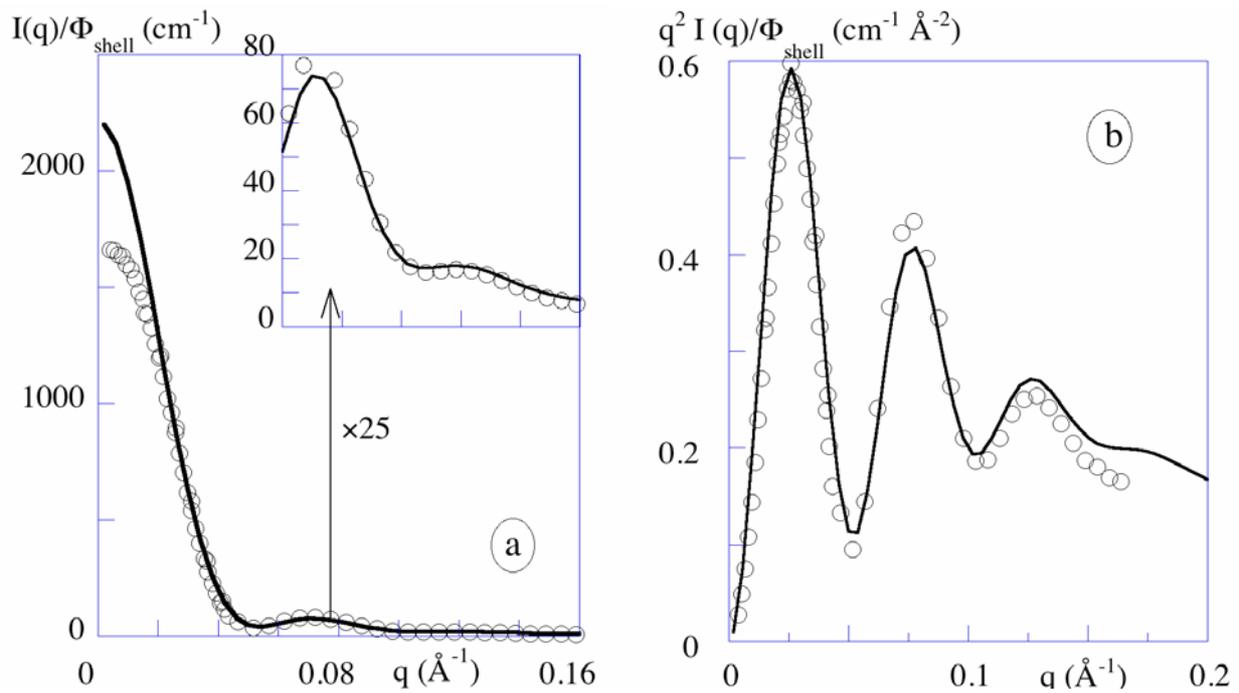

Figure 2 SANS spectra for the bare microemulsion $\Phi_{shell} = 0.0153$ ; circles = experimental data , line = computed form factor [a] The intensity normalized to unit shell volum fraction versus $q$ . [b] Plot of $q^2 I(q)/\Phi_{shell}$ versus $q$ to amplify the form factor oscillations. (See text for further discussion)

In figure 2 , the spectra for the bare microemulsion is shown, the oscillations of the form factor, already perceptible in the representation $I(q)$ versus $q$, are amplified in the representation of $q^2 I(q)$ versus $q$. The solid line is the form factor $AP(q)$ computed with eq (15). It is a good fit to the experimental spectra at high q values where $S(q) \approx 1$, while in the low $q$ range evidence of a repulsive interaction is found: $S(q) < 1$. In figure 3 a typical spectra of the microemulsion decorated with the copolymer is shown. Qualitatively one can note a large discrepancy between the experimental spectra at low $q$ values and the computed form factor : the corona of PEO introduces a large steric repulsive interaction between the droplets evidenced by the correlation peak. At large $q$ values a good fit is achieved, the droplets are well described by spherical shells but the shift of the minima and maxima to larger $q$'s indicates that the droplets become smaller. In this paper, we focus on the form factor its evolution allows for the determination of $\bar{R}_m \delta$ and $\Delta R$; a description of the increasing repulsive interaction brought about by the PEO corona will be given in a forthcoming paper.

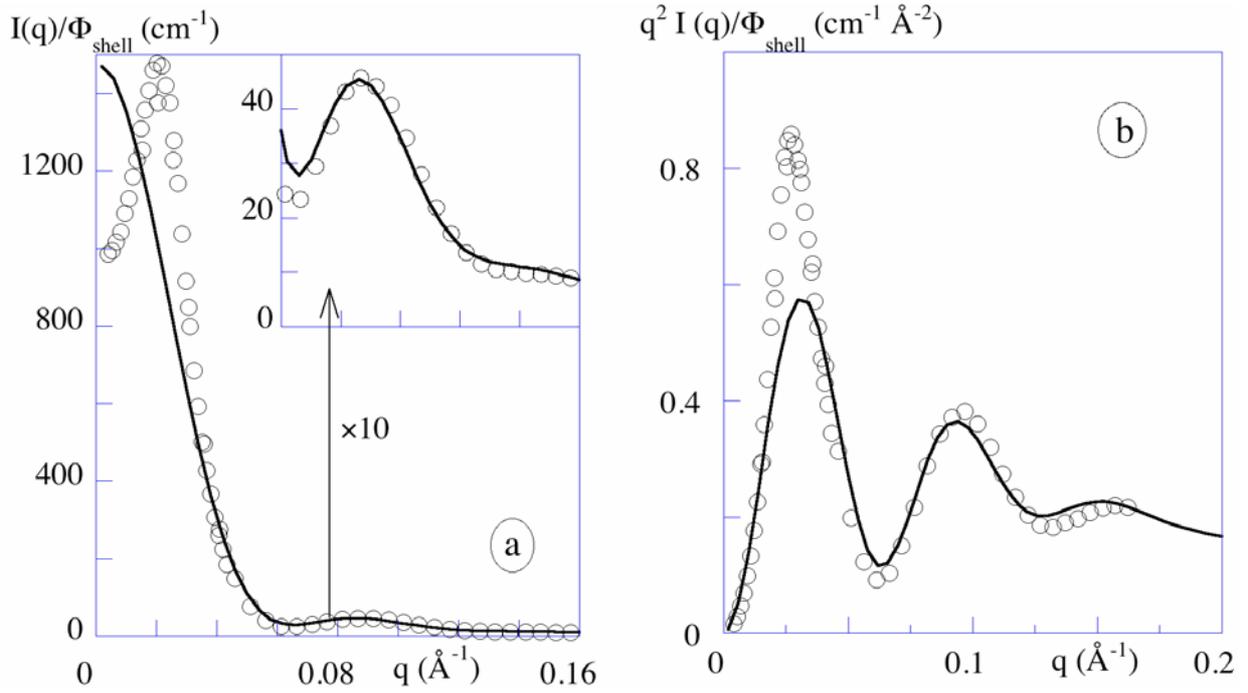

Figure 3 SANS spectra for the microemulsion $\Phi_{shell} = 0.0153$ decorated by ~ 27 Polymers per droplet ($\sigma = 6.85 \times 10^{-4}$ Å$^{-2}$); circles = experimental data, line = computed form factor [a] The intensity I(q) normalized to unit shell volum fraction versus $q$ [b] Plot of $q^2 I(q)/\Phi_{shell}$ versus $q$ to amplify the form factor oscillations. (See text for further discussion).

**Results and Discussion**

From the spectra measured for the various samples (bare to highly decorated microemulsion) we obtain the mean radius $\overline{R}_m$, the shell thickness $\delta$ and the standard deviation $\Delta R$ as describe above.

In the bare microemulsion the shell of the droplets contains the CPCl and octanol molecules, in the decorated microemulsion, some surfactant molecules are replaced by the copolymer in order to keep a practically constant shell volume fraction as indicated above. For each sample we can calculate the number $N_{drops}$ of droplets per unit volume : $N_{drops} = \phi_{shell}/v_{shell}$ ($v_{shell}$ is given by eq 14) and thus the total surface of the droplets covered by the surfactant monolayer including, in the decorated microemulsion, the C12 chains of PEO-C12. The decoration of the microemulsion droplets is characterized by $\sigma$ the number of grafted chains per unit aera of the monolayer estimated on the outer surface of the droplets, we discuss the choice of this particular surface below.

$$\sigma = \frac{v_{shell}}{\phi_{shell}} \frac{1}{4\pi(R_m + \delta/2)^2} \times \frac{c_{PEO-C12}}{M_{PEO-C12}/N_A}$$ with $c_{PEO-C12}$ and $M_{PEO-C12}$ respectively the

weight concentration and the molar mass of copolymer and $N_A$ Avogadro's number.

With the radius of giration of the PEO-chain estimated from the relationship[36, 37]: $R_g = 0.107 M^{0.63} = 23.5 \text{Å}$ we calculate $\sigma* = 5.8 \; 10^{-4} \; \text{Å}^{-2}$, the overlap coverage which is the threshold value for the brush regime for a flat surface[22]. Note however, that for a sharply curved sphere ($R_g \approx R$), the real threshold value between the mushroom and the brush regimes should be shifted to a significantly higher value than for the flat case. It means than in our experiments, where $\sigma$ is at most equal to $2.2 \times \sigma*$, the brush regime is never reached.

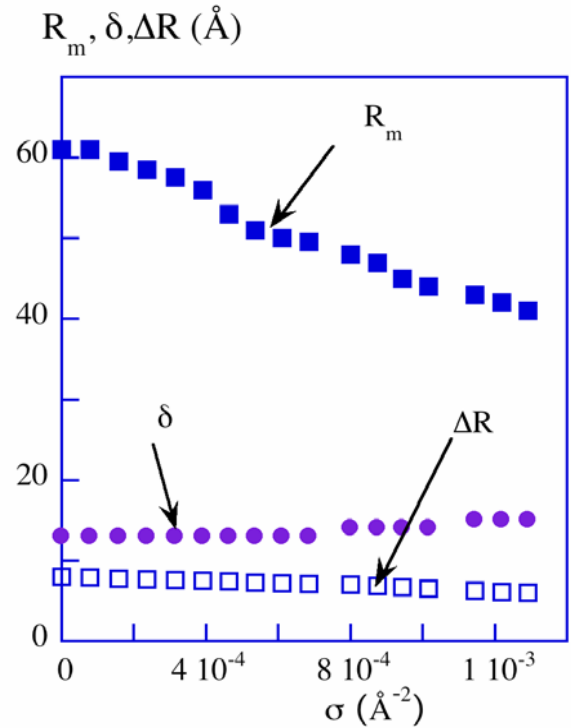

Figure 4 The values for the mean radius $\overline{R_m}$, the thickness $\delta$ and the standard deviation $\Delta R$ obtained from the fit of the SANS spectra to a shell form factor are plotted as a function of $\sigma$. $\overline{R_m}$ decreases regularly from 60 to 40 Å.

The evolution of $\overline{R_m}, \delta$ and $\Delta R$ with $\sigma$ is illustrated in figure 4, the mean radius decreases monotically and the shell thickness $\delta$ remains practically constant. With the geometrical parameters and the number of droplets, an estimation of the volume fraction of decane in the microemulsion droplets indicates a decrease from 0.0158 to 0.007, in good agreement with the qualitative observation of the rejected decane in the samples. In the following discussion, we assume that the totality of the surfactant CPCl and the cosurfactant octanol remains with the microemulsion.

From these results, eq. 2 to 4 and $\phi = N_{drops} \frac{4\pi}{3}(R_m + \frac{\delta}{2})^3$ we derive the values of $\tilde{R}_0 = \tilde{C}_0^{-1}$ and $\tilde{\kappa}$ applying the equations at the outer surface of the droplet: $R = R_m + \delta/2$ and $p = \Delta R / (R_m + \delta/2)$. At this point a question must be raised, eq. 2 to 4 are derived assuming no coupling between the curvature and the area per surfactant molecule, this arises only at the neutral surface of the film where the area per surfactant molecule is independant of the degree of bending and equal to the area in the flat film [1]. We implicitly assumed above, that this neutral surface (or surface of no tension) is the outer surface of the surfactant film. The number of aliphatic chains in the film varies by less than a few ‰ in the samples studied within the plausible assumption that all the surfactants remain in the film i.e. pure decane is rejected from the samples. We can then compute the surface per aliphatic chain on the outer surface of the droplets and we find that it shows very small variations (± 2 Å$^2$) around 42 Å$^2$ while the surface per aliphatic chain decreases by roughly 14 % at the middle plane of the film and by more than 25% at the inner plane of the film. Thus the area per molecule is constant at the outer surface of the droplets justifying a posteriori our assumption ; in fact for such a surfactant film one generally assumes the neutral surface to be the separation surface between the polar heads and the aliphatic chains which is here (considering the nature of the surfactants) very close (2 or 3 Å) to the outer surface.

The obtained values are plotted in figure 5 as a function of $\sigma$ together with the calculated values $R_{0,theo} = \tilde{C}_{0,\mu}^{-1}$ and $\tilde{\kappa}_\mu$, given respectively by equations 12-b and 12-a of the model. Experimentally, $\tilde{\kappa}$ remains constant within experimental errors and equal 1.5 ± 0.2 kT and the optimum radius decreases with increasing decoration. Both the decrease of $\tilde{R}_0$ and the constant value of $\tilde{\kappa}$ are correctly predicted by the calculations  In these calculations we set the polymerization index $N$=113 (the mean number of EO monomers in the chain). $\tilde{\kappa}_{bare}$ = 1.6 kT and $\tilde{C}_{0,bare}^{-1}$= 76 Å are the experimental values measured for the bare microemulsion. The monomer length $a$ is the only fitting parameter of our model. The best fitting value is $a$ = 3. 5 Å. As discussed in the theoretical section, choosing $a$, is a mean to adapt the model to real chains , even so it is derived for Gaussian chains. Increasing $a$ will lead to a sharper decrease of $R_{0,theo}$ with $\sigma$, but the variations of $R_{0,theo}$ with $\sigma$ will remain comparable to the experiments for meaningful values of $a$ ( 2 Å < $a$ < 6Å). On the other hand, the variations of $\tilde{\kappa}_\mu$ with $\sigma$ remain less than 0.05 $kT$ whatever the value of $a$.. This would not be the case if we choose for comparison  with the experiments, $\tilde{\kappa}_\sigma$ rather than $\tilde{\kappa}_\mu$. This result illustrates the contribution of the free diffusion of polymer chains to the bending properties of the surface. Notice that the same quasi invariance of the bending modulus upon the chain coverage has already been observed in hairy vesicules[5].

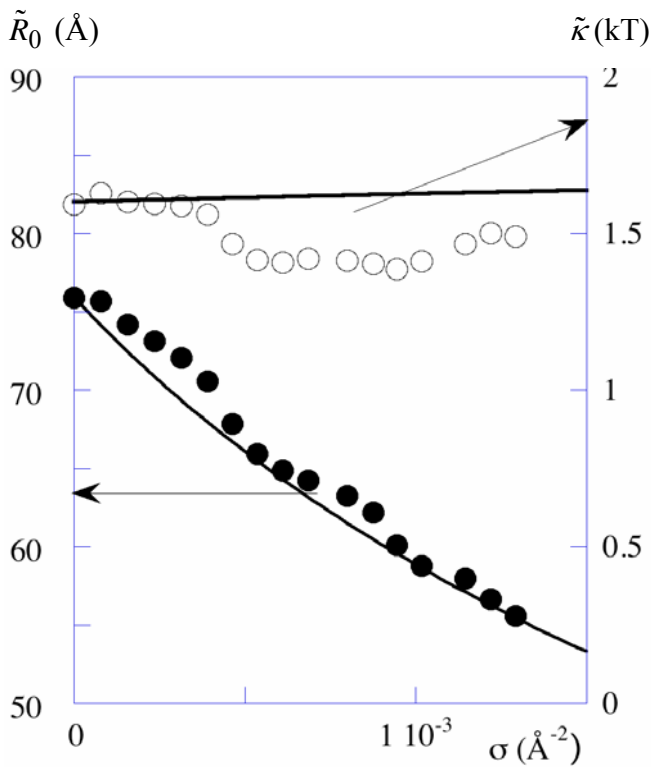

Figure 5

The elastic properties of the surfactant film as a function of its decoration by the anchoring of PEO-C12 measured by $\sigma$ the number of PEO-C12 per Å$^2$. The coefficient of rigidity $\tilde{\kappa}$ (open circles) is found almost constant (within experimental errors) and equal to $1.5 \pm 0.2$ kT. The optimum radius of curvature (full dots) $\tilde{R}_0 = \tilde{C}_0^{-1}$ decreases slowly and monotically. The calculated values are the lines. (see text for further discussion)

## Acknowledgments

This work was conducted as part of the scientific program of the Network of Excellence "Soft Matter Composites : an approach to nanoscale functional materials" supported by the European Commission.

We thank Raymond Aznar for synthetizing the copolymer used in this work.

We thank Loic Auvray and Didier Lairez for their help during the SANS experiments performed on line PACE at Laboratoire Léon Brillouin -CEA-CNRS.

# References

[ 1] Safran S A, Curvature elasticity of thin films, 1999, Adv. Phys., **48**, 395

[ 2] Sottmann T, Solubilization efficiency boosting by amphiphilic block co-polymers in microemulsions, 2002, Curr. Opin. Colloid Interface Sci., **7**, 57

[ 3] Lipowsky R, Flexible membranes with anchored polymers, 1997, Colloid Surf. A-Physicochem. Eng. Asp., **128**, 255

[ 4] Castro-Roman F, Porte G and Ligoure C, Smectic Phase of fluid membranes decorated by amphiphilc copolymers, 2001, Langmuir, **17**, 5045

[ 5] Joannic R, Auvray L and Lasic D D, Monodisperse vesicles stabilized by grafted polymers, , 1997, Phys. Rev. Lett., **78**, 3402

[ 6] Prince L M, Microemulsions: Theory and Practice, 1977, Academic Press, New York

[ 7] De Gennes P G and Taupin C, Micro-emulsions and the flexibility of oil-water interfaces, 1982, J.Phys.Chem., **86**, 2294

[ 8] Andelman D, Cates M E, Roux D and Safran S, Structure and phase-equilibria of microemulsions, 1987, J. Chem. Phys., **87**, 7229

[ 9] Safran S A and Turkevich L A, Phase-Diagrams for Microemulsions, 1983, Phys. Rev. Lett., **50**, 1930

[ 10] Langevin D, 1991, Adv.Colloid Interf.Sci. , **34**, 583

[ 11] Strey R, Phase behavior and interfacial curvature in water-oil-surfactant systems, 1996, Curr. Opin. Colloid Interface Sci., **1**, 402

[ 12] Gradzielski M and Langevin D, Small-angle neutron scattering experiments on microemulsion droplets: Relation to the bending elasticity of the amphiphilic film, 1996, J. Mol. Struct., **383**, 145

[ 13] Gradzielski M, Langevin D and Farago B, Experimental investigation of the structure of nonionic microemulsions and their relation to the bending elasticity of the amphiphilic film, 1996, Phys. Rev. E**, 53**, 3900

[ 14] Gradzielski M, Langevin D, Sottmann T and Strey R, Droplet microemulsions at the emulsification boundary: The influence of the surfactant structure on the elastic constants of the amphiphilic film., 1997, J. Chem. Phys., **106**, 8232

[ 15] Farago B and Gradzielski M, The effect of the charge density of microemulsion droplets on the bending elasticity of their amphiphilic film, 2001, J. Chem. Phys., **114**, 10105

[ 16] Schulman J H, Stoeckenius W and Prince L M, Mechanism of Formation and Structure of Micro Emulsions by Electron Microscopy, 1959, J.Phys.Chem., **63**, 1677

[ 17] Safran S A, Saddle-Splay Modulus and the Stability of Spherical Microemulsions, 1991, Phys. Rev. A, **43**, 2903

[ 18] Lipowsky R, Bending of Membranes by Anchored Polymers, 1995, Europhys. Lett., **30**, 197

[ 19] Hiergeist C and Lipowsky R, Elastic properties of polymer-decorated membranes, 1996, J. Phys. II (France), **6,** 1465

[ 20] Marques C M and Fournier J E L, Deviatoric spontaneous curvature of lipid membranes induced by Siamese macromolecular cosurfactants, , 1996, Europhys. Lett., **35**, 361

[ 21] Gomper G, Richter D and Strey R, Amphiphilic block copolymers in oil-water-surfactant mixtures: efficiency boosting, structure, phase behaviour and mechanism, 2001, J.Phys. Cond. Matt., **13**, 9055

[ 22] De Gennes P G, Conformations of Polymers attached to an interface, 1980, Macromolecules, **13**, 1069

[ 23] Bickel T and Marques C M, Scale dependent rigidity of polymers grafted to soft interfaces, 2002, Eur. Phys. J. E, **9**, 349

[ 24] Helfrich W, Elastic properties of lipid bilayers: theory and possible experiments., 1973, Z. Naturforsch.C, **28**, 693

[ 25] Safran S A and Tlusty T, Curvature elasticity models of microemulsions, 1996, Ber. Bunsen-Ges. Phys. Chem. Chem. Phys., **100**, 252

[ 26] Gradzielski M, Bending constants of surfactant layers, 1998, Curr. Opin. Colloid Interface Sci., **3**, 478

[ 27] Milner S T and Safran S A, Dynamic Fluctuations of Droplet Microemulsions and Vesicles, 1987, Phys. Rev. A, **36**, 4371

[ 28] Safran S A, Fluctuations of Spherical Microemulsions, 1983, J. Chem. Phys., 78, 2073

[ 29] Porte G and Ligoure C, Mixed amphiphilic bilayers :bending elasticity and formation of vesicles, 1995, J. Chem. Phys., **102**, 4290

[ 30] Auth T and Gompper G, Self-avoiding linear and star polymers anchored to membranes , 2003, Phys.Rev.E, **68**, 051801

[ 31] Bickel T, Jeppesen C and Marques C M, Local entropic effects of polymers grafted to soft interfaces, 2001, Eur. Phys. J. E, **4**, 33

[ 32] Hartmann P, Collet A and Viguier M, Synthesis and characterization of model fluoroacylated poly(ethylene oxide), 1999, J. Fluor. Chem., **95**, 145

[ 33] Filali M, Aznar R, Svenson M, Porte G and Appell J, Swollen micelles plus hydrophobically modified hydrosoluble polymers in aqueous solutions: Decoration versus bridging. A small angle neutron scattering study, 1999, J. Phys. Chem. B, **103**, 7293

[ 34] Lairez D, Résolution d'un spectromètre de diffusion de neutrons aux petits angles, 1999, J. Phys IV (France), **9**, 1

[ 35] A Gaussian distribution is equivalent to the Schultz distribution as long as the polydispersity index remains small ( less than 0.3)

[ 36] Cabane B, Small Angle Scattering Methods, 1987, Surfactant solutions: new methods of investigation, 57

[ 37] Benkhira A, Franta E, Rawiso M and Francois J, Conformation of Poly(1,3-Dioxolane) in Dilute and Semidilute Aqueous-Solutions, 1994, Macromolecules, **27**, 3963

**Figure Captions**

<u>Figure 1</u> A microemulsion droplet with one PEO-C12 polymer anchored via its C12 aliphatic end in the surfactant shell. Water and decane are deuterated, neutron scattering is essentially due to the surfactant shell, the PEO chains are very few and do not contribute significantly to the scattering.. In the model, the shell is characterized by its mean radius $R_m = (R_1 + R_2)/2$ and its thickness $\delta = R_2 - R_1$ For the bare microémulsion $R_m \approx 61 \text{Å}$ and $\delta \approx 13 \text{Å}$; the PEO chain has a giration radius $R_g \approx 23.5 \text{Å}$

<u>Figure 2</u> SANS spectra for the bare microemulsion $\Phi_{shell} = 0.0153$; circles = experimental data, line = computed form factor [a] The intensity normalized to unit shell volum fraction versus $q$. [b] Plot of $q^2 I(q)/\Phi_{shell}$ versus $q$ to amplify the form factor oscillations. (See text for further discussion)

<u>Figure 3</u> SANS spectra for the microemulsion $\Phi_{shell} = 0.0153$ decorated by ~ 27 Polymers per droplet ($\sigma = 6.85 \times 10^{-4}$ Å$^{-2}$); circles = experimental data, line = computed form factor [a] The intensity I(q) normalized to unit shell volum fraction versus $q$ [b] Plot of $q^2 I(q)/\Phi_{shell}$ versus $q$ to amplify the form factor oscillations. (See text for further discussion).

<u>Figure 4</u> The values for the mean radius $\overline{R}_m$, the thickness $\delta$ and the standard deviation $\Delta R$ obtained from the fit of the SANS spectra to a shell form factor are plotted as a function of $\sigma$. $\overline{R}_m$ decreases regularly from 60 to 40 Å.

<u>Figure 5</u>

The elastic properties of the surfactant film as a function of its decoration by the anchoring of PEO-C12 measured by $\sigma$ the number of PEO-C12 per Å$^2$. The coefficient of rigidity $\tilde{\kappa}$ (open circles) is found almost constant (within experimental errors) and equal to 1.5 ± 0.2 kT. The optimum radius of curvature (full dots) $\tilde{R}_0 = \tilde{C}_0^{-1}$ decreases slowly and monotically. The calculated values are the lines. (see text for further discussion)